\documentclass[aps,pre,twocolumn,twoside, amsmath,amssymb,aps,bbm,floatfix,showpacs]{revtex4-1}

\usepackage{xcolor}
\usepackage{graphicx}
\usepackage{subfigure}
\usepackage{dcolumn}
\usepackage{bm}
\usepackage{framed}
\usepackage{hyperref}
\usepackage{dsfont}
\usepackage[normalem]{ulem}
\usepackage{nicefrac}

\usepackage{nicefrac}
\usepackage{braket}

\newcommand{\be}{\begin{equation}}
\newcommand{\ee}{\end{equation}}
\newcommand{\bea}{\begin{eqnarray}}
\newcommand{\eea}{\end{eqnarray}}
\newcommand{\la}{\left\langle}
\newcommand{\ra}{\right\rangle}

\begin{document}

\title{Efficiency fluctuations of a quantum Otto engine}
\author{Tobias Denzler}
\author{Eric Lutz}
 \affiliation{Institute for Theoretical Physics I, University of Stuttgart, D-70550 Stuttgart, Germany}

\begin{abstract}

We derive the probability distribution of the efficiency of a quantum Otto engine. We explicitly compute the quantum efficiency statistics for an analytically solvable two-level engine. We analyze the occurrence of values of the stochastic efficiency above unity, in particular at infinity,  in the nonadiabatic regime and further determine mean and variance  in the case of adiabatic driving. We finally investigate the classical-to-quantum transition as the temperature is lowered.
\end{abstract}

\maketitle

Efficiency is a key performance measure of thermal machines. For  heat engines  that cyclically convert heat into useful work, it is defined as the ratio of work output and heat input \cite{cen01}. For macroscopic systems consisting of a huge number of degrees of freedom, heat, work and, consequently, efficiency are deterministic quantities. The second law of thermodynamics imposes an upper bound to the efficiency of any heat engine given by the familiar Carnot expression, $\eta_\text{carnot}= 1- T_1/T_2$, where $T_{1,2}$ are the respective temperatures of the cold and hot heat reservoirs \cite{cen01}. By contrast, these variables become random at the microscopic scale owing to the presence of nonnegligible thermal fluctuations  \cite{sei12,sek98} and additional  quantum  fluctuations  at low enough temperatures  \cite{esp09,cam11}. A central question is then to determine their probability distributions in order to assess their stochastic  properties.

For classical heat engines, efficiency fluctuations have been theoretically investigated in a number of recent studies \cite{ver14,ver14a,gin14,pol15,pro15,pro15a,pro16,vro16,par16,pro17}. In particular, the stochastic efficiency of a Carnot engine has been shown to admit values larger than the Carnot bound and, remarkably, to be the least likely in the long-time limit \cite{ver14}. These predictions have been experimentally verified in a stochastic Carnot engine based on  an optically trapped colloidal particle \cite{mar15}. In the quantum regime,    work distributions of  driven  oscillators have been  analyzed both theoretically \cite{def08,tal09,def10} and experimentally  using a trapped ion \cite{an15}. The quantum work statistics of driven two-level systems  has been similarly computed analytically  \cite{sol13,hek13} and determined experimentally in  NMR  \cite{bat14} and cold-atom \cite{cer17} setups. On the other hand, the quantum heat statistics has been calculated theoretically for harmonic oscillators \cite{den18} as well as for two-level models \cite{gas14,pon15},  and reconstructed experimentally in the latter case \cite{pet18}.  However, to our knowledge, the effects of quantum fluctuations on the efficiency of a heat engine have not been examined so far.

In this paper, we analytically evaluate the efficiency statistics of a quantum Otto engine, a paradigmatic model of quantum thermal machines \cite{kos17}. The quantum Otto cycle, a generalization of the ordinary four-stroke motor, has been extensively studied in the past thirty years \cite{kos84,gev92,fel00,kie04,hen07,scu02,lin03,rez06,aba12,wat17}. The experimental realization of a quantum Otto spin  engine in a NMR system has  been reported lately \cite{pet18}. In the following, we concretely determine the respective work and heat probability densities of the different branches  of the engine cycle within the usual two-projective measurement scheme \cite{tal07}. We use these distributions to derive a general formula for the quantum efficiency statistics that explicitly depends on the time evolution operators of expansion and compression steps. We treat in detail an analytically solvable model of a two-level engine and obtain an explicit expression for the quantum efficiency distribution. We discuss the appearance of values of the stochastic efficiency above unity in the nonadiabatic regime and, in particular, at infinity when no heat is absorbed although non-zero work is produced. This peculiar behavior stems from the discrete quantum nature of the engine. We further determine mean and variance of the efficiency in the case of adiabatic driving. We concretely investigate their evolution from a regime dominated by thermal fluctuations at high temperatures to a domain characterized by quantum fluctuations at low temperatures. We finally demonstrate that the average quantum efficiency is always smaller than the conventional  thermodynamic efficiency owing to the presence of positive correlations between stochastic efficiency and absorbed heat.

\textit{Quantum Otto engine.} We consider a generic quantum system with a time-dependent Hamiltonian $H_t$ as the working fluid of a quantum Otto engine. The system is initially thermalized at $t=0$ by weakly coupling it to a cold heat reservoir at inverse temperature $\beta_1$.
The Otto  cycle consists of the following four steps  (Fig.~1): (1) Unitary expansion (AB) during which the Hamiltonian is changed from $H_0$ to $H_{\tau_1}$ in a time $\tau_1$, consuming an amount of work $W_1$, (2) Hot isochore  (BC) during which the system is put into contact with a heat bath at inverse temperature $\beta_2$ to absorb heat $Q_2$ in a  time $\tau_2$, (3) Unitary compression (CD) that drives the isolated system  from $H_{\tau_1}$ back to $H_0$ in a time $\tau_3$, producing  an amount of work $W_3$, and (4) Cold isochore  (DA), which closes  the cycle  by weakly coupling the system to the cold bath at inverse temperature $\beta_1$, thus releasing heat $Q_4$ in a time $\tau_4$.  We further  assume that heating and cooling times, $\tau_{2,4}$, are longer than the relaxation time of the system, so that  thermalization is achieved after each isochore.   

\begin{figure}[t]
  \includegraphics[width=0.33\textwidth, clip, trim=0cm 0cm 0cm 0cm]{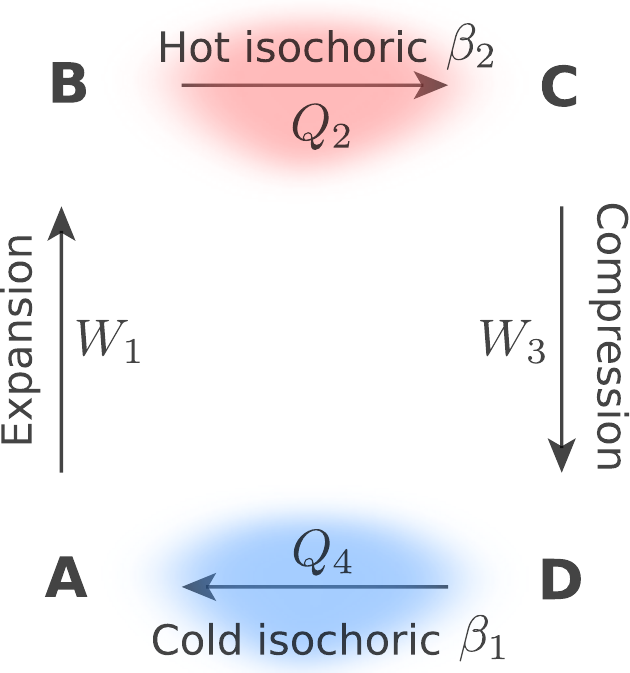}
\caption{Quantum Otto cycle consisting of unitary expansion (AB), isochoric heating at inverse temperature $\beta_2$ (BC), unitary compression (CD) and isochoric cooling at inverse temperature $\beta_1$ (DA). The  Otto engine absorbs heat $Q_2$ (releases heat $Q_4$) and produces the positive work $-(W_1+W_3)$.}\label{fig:1}
\end{figure}

Quantum work and heat distributions are commonly determined with the help of the so-called two-projective-measurement method \cite{tal07} or variants thereof \cite{def08,tal09,def10,an15,sol13,hek13,bat14,cer17,den18,gas14,pon15}. In this approach, energy changes of a quantum system during single realizations  of a process are associated with the difference of  eigenvalues obtained though projective energy measurements at the beginning and at the end of the process. In the quantum Otto cycle, work is performed during the unitary expansion and compression steps, while heat is exchanged during the nonunitary heating and cooling stages. We obtain the corresponding distributions by applying the two-projective-measurement scheme to the respective expansion, hot isochore and compression branches. The probability distribution of the expansion work $W_1$ is accordingly \cite{tal07},
\begin{equation}\label{eq:W1}
	P(W_1)=\sum_{n,m} \delta \left[W_1 - (E_m^\tau- E_n^0)\right] P_{n \rightarrow m}^\tau P_n^0(\beta_1),
\end{equation}
where   $E_n^0$ and $E_m^\tau$ are the respective energy eigenvalues at the beginning and at the end of the expansion step,  $P_n^0(\beta_1)= \exp({-\beta_1 E_n^0})/Z^0$ is the initial thermal occupation probability with partition function $Z^0$ and $P_{n \rightarrow m}^\tau= |\bra{n}U_\text{exp}\ket{m}|^2$ the transition probability from eigenstate $\ket{n}$ to $\ket{m}$. The corresponding unitary time evolution operator is denoted by $U_\text{exp}$. The  occupation probability $P_n^0(\beta_1)$ embodies the influence of thermal fluctuations, whereas the transition probability $P_{n \rightarrow m}^\tau$ accounts for the effects of  quantum fluctuations.

Similarly, the probability density of the heat $Q_2$ during the ensuing hot isochore, given the expansion work $W_1$, is equal to the conditional distribution \cite{jar04},
  \begin{equation}\label{eq:Q2}
	P(Q_2|W_1)=\sum_{k,l} \delta \left[Q_2 -(E_l^{\tau} - E_k^\tau) \right]P_{k \rightarrow l}^{\tau_2} P_k^\tau,
\end{equation}
where the occupation probability at time $\tau$ is $P_k^\tau = \delta_{km}$ when the system is in eigenstate $\ket{m}$ after the second projective energy measurement.  Noting that the state of the system is thermal with inverse temperature $\beta_2$ at the end of the isochore, we further have $P_{k \rightarrow l}^{\tau_2} = P_l^{\tau_2}(\beta_2)=\exp({-\beta_2 E_l^{\tau}})/Z^{\tau}$, with the partition function $Z^{\tau}$.

The quantum work distribution for compression, given the expansion work $W_1$ and the heat $Q_2$, is furthermore,
\begin{equation}\label{eq:W3}
	\!\!\!P(W_3|W_1,Q_2)\!=\! \sum_{i,j} \delta\! \left[ W_3 - (E_j^0 - E_i^\tau) \right] \!P_{i \rightarrow j}^\tau P_i^{\tau+\tau_2},
\end{equation}
with the  occupation probability $P_i^{\tau+\tau_2} = \delta_{il}$ when the system is in eigenstate $\ket{l}$ after the third projective energy measurement. The transition probability $P_{i \rightarrow j}^\tau=|\bra{i}U_\text{com}\ket{j}|^2$ is fully specified by the unitary time evolution operator for compression $U_\text{com}$. 

The joint  probability of having certain values of $W_3$, $Q_2$ and $W_1$ during a cycle of the Otto engine readily follows from the chain rule for conditional probabilities \cite{pap91}, $P(W_3,Q_2,W_1) = P(W_3|Q_2,W_1)P(Q_2|W_1)P(W_1)$. Using Eqs.~\eqref{eq:W1}, \eqref{eq:Q2} and \eqref{eq:W3}, it is explicitly given by,
\begin{eqnarray}\label{eq:p_tot}
P(W_1,Q_2,W_3)& =& \sum_{n,m,k,l} \delta \left[W_1 - (E_m^\tau- E_n^0)\right] \nonumber \\
&\times& \delta \left[Q_2 -(E_k^\tau - E_m^\tau) \right] \delta \left[ W_3 - (E_l^0 - E_k^\tau) \right]	\nonumber \\
&\times& |\bra{n} U_\text{exp}\ket{m}|^2 |\bra{k} U_\text{com}\ket{l}|^2 \nonumber\\
&\times&  \frac{e^{-\beta_1 E_n^0} e^{-\beta_2 E_k^\tau}}{Z^0 Z^\tau}.
\end{eqnarray}
Equation \eqref{eq:p_tot} is the essential quantity needed to determine the quantum efficiency statistics.

\textit{Quantum efficiency distribution.}
The stochastic efficiency of the quantum Otto engine  is defined as,
\begin{equation}\label{eq:def_eta}
	\eta = -\frac{W_1+W_3}{Q_2}.
\end{equation}
It should not be confused with the thermodynamic efficiency which is defined in terms of the averaged values of work and heat, $\eta_\text{th}= -(\la W_1\ra + \la W_3\ra)/ \la Q_2\ra$.
The  probability distribution $P(\eta)$ is obtained by integrating over all  possible values of $W_3$,  $Q_2$ and $W_1$ via,
\begin{equation}
	P(\eta) =\! \int dW_3 dQ_2 dW_1\,P(W_1,Q_2,W_3)\,\delta \left( \eta +  \frac{W_1 + W_3}{Q_2}\right)
\end{equation}
Using Eq.~\eqref{eq:p_tot} and the properties of the  delta, we find,
\begin{eqnarray}\label{eq:p_eta}
P(\eta) &=& \sum_{n,m,k,l} \delta \left(\eta  + \frac{E_m^\tau -E_n^0+E_l^0-E_k^\tau}{E_k^\tau - E_m^\tau} \right) \\
		& \times&  \frac{e^{-\beta_1 E_n^0} e^{-\beta_2 E_k^\tau}}{Z^0 Z^\tau} |\bra{m}U_\text{exp}\ket{n}|^2 |\bra{k}U_\text{com}\ket{l}|^2. \nonumber
\end{eqnarray}
Expression \eqref{eq:p_eta} is our main result. It shows that the efficiency statistics of the quantum Otto engine is fully determined by the unitary time evolution operators for  expansion and compression, $U_\text{exp}$ and $U_\text{com}$, and by the inverse temperatures, $\beta_1$ and $\beta_2$, of the two reservoirs,  when complete thermalization is assumed.

\textit{Example of a spin heat engine.} Formula \eqref{eq:p_eta} is valid for any working fluid. As an illustration, we now investigate the fluctuating properties of the stochastic efficiency \eqref{eq:def_eta} for an analytical solvable two-level heat engine. Compression and expansion are implemented by driving a spin-$1/2$ with  a constant magnetic field with strength $\omega/2$ along the $z$-axis and a rotating magnetic field with varying strength $\gamma(t)$ in the ($x$-$y$)-plane. This driving  changes both the eigenenergies and the occupation probabilities of the system and could be realized in a NMR setup \cite{pet18}. The expansion Hamiltonian reads,
\begin{equation}
H_\text{exp}(t)= \gamma(t) \left( \cos\omega t \,\sigma_x + \sin\omega t \,\sigma_y \right) + \frac{\omega}{2} \sigma_z,
\end{equation}
where $\sigma_i$, $i=(x,y,z)$, are the usual Pauli operators.
The rotation frequency is chosen to be $\omega=\pi/2\tau$ to ensure   a complete rotation from the $x$-axis to the $y$-axis during the expansion step of duration $\tau$.
The amplitude of the rotating  field, $\gamma(t)= \gamma_1 \left(1-t/\tau\right)+ \gamma_2  \left(t/\tau\right)$, is increased from $\gamma_1$ at time zero to $\gamma_2$ at time $\tau$. This driving leads to a widening of  the energy spacing of the two-level system from 
$\nu^0= \sqrt{4\gamma(0)^2+\omega^2}/2$ to $\nu^\tau=\sqrt{4\gamma(\tau)^2+\omega^2}/2$. For simplicity, we will take expansion and compression times to be equal,  $\tau_1=\tau_3=\tau$.
The compression stroke is then simply obtained from the time reversed process, $H_\text{com}(t)=-H_\text{exp}(\tau-t)$. The corresponding expansion time evolution operator reads \cite{sup},
\begin{equation}
\label{9}
	U_\text{exp} = 
	\begin{pmatrix}
		e^{-i \omega t/2}\cos I & ie^{-i \omega t/2}\sin I\\
		ie^{i \omega t/2}\sin I & e^{i \omega t/2}\cos I
    \end{pmatrix},
\end{equation}
where $I=-\int_0^t dt' \gamma(t')$ is the integral over the increasing strength of the rotating magnetic field. The operator $U_\text{com}$ follows from $U_\text{exp}$ by replacing $t$ with $\tau -t$.  The probability of  no level transition during expansion or compression steps may be inferred from \eqref{9} as,
\begin{equation}
	u= u_\text{exp}=u_\text{com}  = \cos^2 I.
\end{equation}
The two are identical since  $I$ is the same for both cases  apart from a minus sign.  The probability of a level transition during either driving phases is accordingly $v = 1-u$.

\begin{figure}[t]
  \includegraphics[width=0.49\textwidth, clip, trim=0cm 0cm 0cm 0cm]{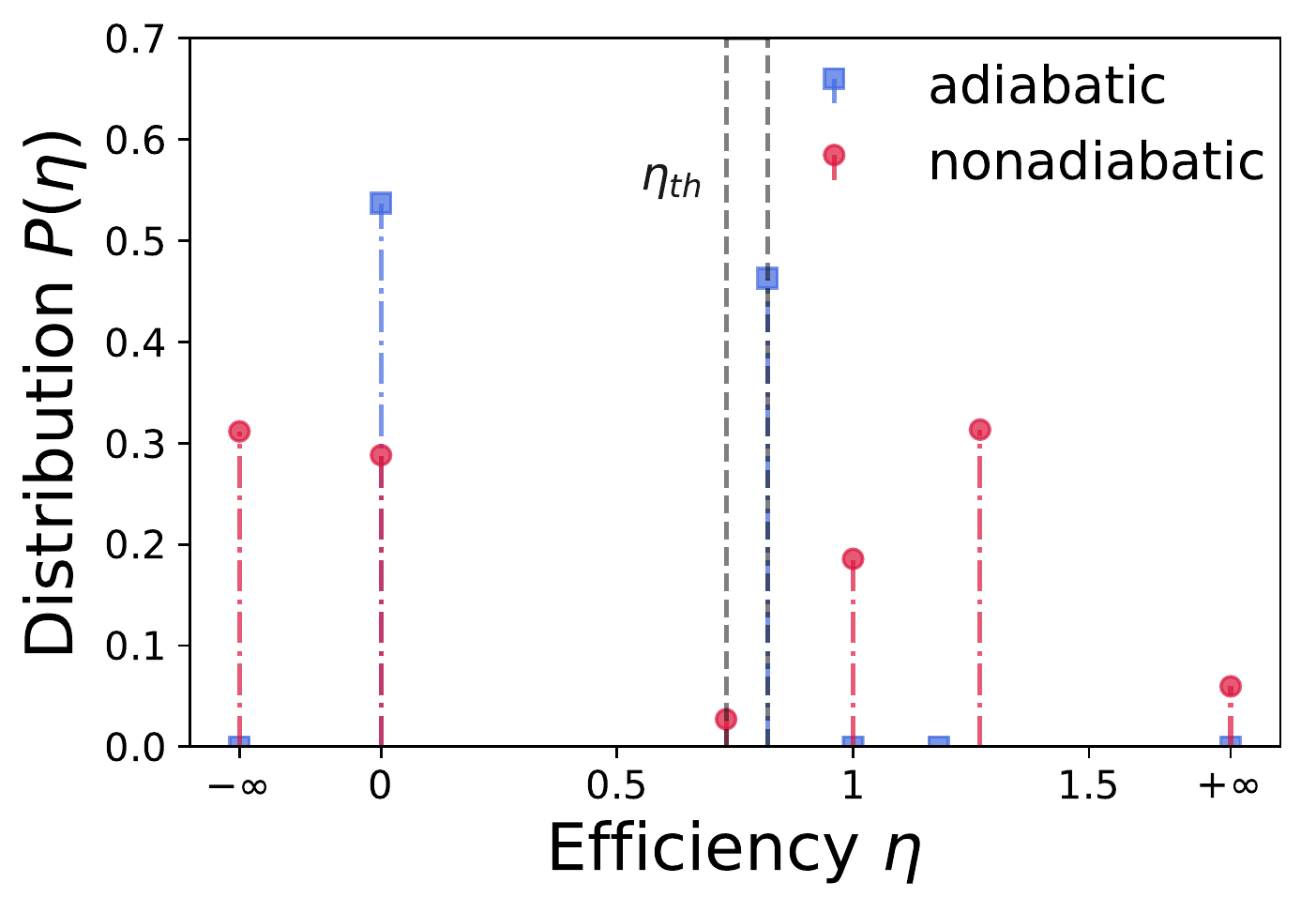}
\caption{Discrete probability distribution $P(\eta)$, Eq.~(17), of the quantum stochastic efficiency (5) for both  adiabatic (blue squares) and nonadiabatic (red dots) driving. We observe the appearance of peaks above unit efficiency, in particular at infinity when the absorbed heat vanishes, in the nonadiabatic regime. They are induced by quantum fluctuations. Parameters are $\beta_1=2$, $\beta_2=0.1$ and  $\gamma_1=0.5$, $\gamma_2=3$. Adiabatic and nonadiabatic driving times are $\tau_\text{ad}=7.18$ and  $\tau_\text{nad}=2.39$.}\label{fig:2}
\end{figure}

In order to operate as an engine, the mean heat absorbed during  (BC) should be positive, $\la{Q_2}\ra>0$, as well as the total mean work output during the cycle, $-(\la{W_1}\ra+\la{W_3}\ra)>0$. Specifically, we have \cite{sup},
\begin{eqnarray}
		\langle W_1 \rangle &=& \left(\nu^\tau A^* +\nu^0 \right) \tanh(\beta_1 \nu^0),	 \\
				\langle W_3 \rangle &=&  \left(\nu^0 A^* + \nu^\tau \right) \tanh(\beta_2 \nu^\tau), \\
				\langle Q_{2}\rangle &=& - \nu^\tau \left[ \tanh(\beta_2 \nu^\tau) + \tanh(\beta_1 \nu^0) A^* \right],	
\end{eqnarray}
where we have introduced the adiabaticity parameter $A^*= 1 - 2u \in [-1,1]$. For adiabatic driving, when the system remains in the same state ($u=1$),  $A^*=-1$, while $A^*=1$ when a transition occurs with certainty ($v=1$). The value   of $A^*$ depends on the driving protocol $\gamma (t)$ as well as on the driving time $\tau$. The two heat engine conditions then lead to the inequalities,
\begin{eqnarray}
		A^* &\leq& - \frac{\tanh(\beta_2 \nu^\tau) }{\tanh(\beta_1 \nu^0)} \label{eq:bound_1}, \\
		A^* &\leq& - \frac{\nu^0 \tanh(\beta_1 \nu^0) + \nu^\tau \tanh(\beta_2 \nu^\tau)}{\nu^\tau \tanh(\beta_1 \nu^0) + \nu^0 \tanh(\beta_2 \nu^\tau)}.\label{eq:bound_2}
	\end{eqnarray}
Equations \eqref{eq:bound_1} and \eqref{eq:bound_2} impose constraints on the allowed values of the  time $\tau$ for a given protocol $\gamma(t)$. The thermodynamic efficiency $\eta_\text{th}$ is further given by,
\begin{equation}\label{eq:eta_mean_2}
\eta_\text{th}= 1 + \frac{\nu^0}{\nu^\tau} \frac{\tanh(\beta_2 \nu^\tau)A^* + \tanh(\beta_1 \nu^0)}{\tanh(\beta_2 \nu^\tau) + \tanh(\beta_1 \nu^0) A^*}.	
\end{equation}
It reduces to the known adiabatic Otto efficiency, $\eta^\text{ad}_\text{th}= 1 - {\nu^0}/{\nu^\tau}$, in the limit $A^*=-1$, as expected \cite{kos17,kos84,gev92,fel00,kie04,hen07,scu02,lin03,rez06,aba12,wat17}.

\begin{figure}[t]
  \includegraphics[width=0.49\textwidth, clip, trim=0cm 0cm 0cm 0cm]{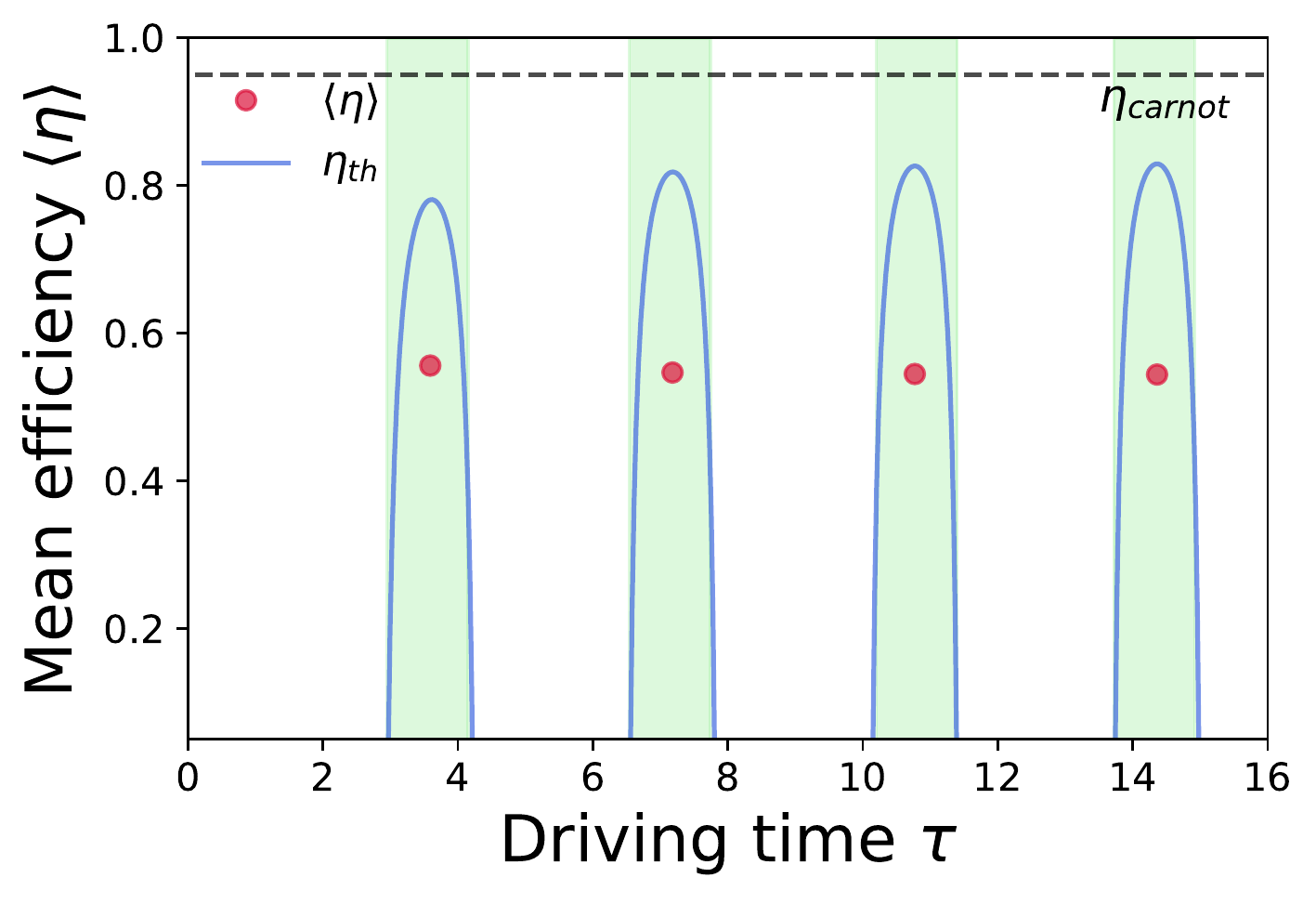}
\caption{Mean adiabatic efficiency $\la \eta\ra$, Eq.~(18) (red dots), and thermodynamic efficiency $\eta_\text{th}$ (blue solid) as a function of the driving time $\tau$. The shaded area, defined by inequalities (14)-(15), represents the regime where the quantum Otto machine operates as a heat engine.  The average adiabatic efficiency is smaller than the corresponding thermodynamic efficiency, $\la \eta\ra< \eta_\text{th}$, due to positive correlations between stochastic efficiency and absorbed heat. Same parameters as in Fig.~2.}\label{fig:3}
\end{figure} 

Using the above results, the quantum efficiency distribution \eqref{eq:p_eta} may be analytically evaluated as,
\begin{eqnarray}
\label{eq:p_eta_2}
			P(\eta)&=& \frac{2}{Z^0 Z^{\tau}} \bigg\lbrace  \left[ u^2 \cosh(\beta_1 \nu^0 + \beta_2 \nu^\tau) \right. \nonumber \\
			&&\left. \left. +v^2 \cosh(\beta_1 \nu^0 - \beta_2 \nu^\tau) \right] \delta \left(\eta  \right)  \right.\nonumber \\
				&&+ v^2 \cosh(\beta_1 \nu^0 + \beta_2 \nu^\tau) \delta \left[\eta - \left(1+ \frac{\nu^0}{\nu^\tau} \right)\right] \nonumber \\
										&&+ \left. u^2 \cosh(\beta_1 \nu^0 - \beta_2 \nu^\tau) \delta \left[\eta - \left(1- \frac{\nu^0}{\nu^\tau} \right) \right] \right.  \nonumber \\
										&&+\left. \cosh(\beta_2 \nu^\tau) \left (e^{\beta_1 \nu^0} \delta [\eta+\infty] + e^{-\beta_1 \nu^0} \delta[\eta-\infty]\right) \right  \rbrace \nonumber	\\			 						&&+ uv  \left[ \delta (\eta -1)  \right] .
\end{eqnarray}
This distribution is normalized to one, as it should. 
We observe that the stochastic efficiency $\eta$ can take six different discrete values as seen in Fig.~\ref{fig:2}. Four are particularly notable: (i) the value at zero is obtained when the produced work $-(W_1+W_3)$ vanishes \cite{sup}, (ii) on the other hand, the value at one corresponds to the case where the eigenvalue at point $A$ is the same as the one at point $D$, implying $W_1+W_3+Q_2=0$, (iii) finally  the values at infinity occur when the heat $Q_2$ is zero. All four follow from the discrete quantum nature of the energy spectrum. The values at infinity are particularly intriguing, since the  efficiency is not defined at these points. 

The efficiency statistics \eqref{eq:p_eta_2}  depends on the driving time $\tau$ (see Fig.~\ref{fig:2}). For adiabatic driving, $u=1$ $  (v=0)$ (blue squares), all values of  the stochastic efficiency are smaller or equal than the adiabatic Otto efficiency $\eta^\text{ad}_\text{th}$, with the largest peak at zero and the second largest at  $\eta^\text{ad}_\text{th}$. The value at infinity does not appear in this case. By contrast, for  nonadiabatic driving, $v> 0$ $  (u<1)$ (red dots), three peaks at and above unit efficiency  are  visible, including the one at infinity. As a result, an average efficiency is not defined. Specifically, Eq.~\eqref{eq:p_eta_2} reveals that the values at one and infinity only disappear when $uv=0$, that is, for certain events. They may thus be regarded as following from quantum indeterminacy.

In the adiabatic case,  $u=1$, the mean efficiency reads,
\begin{equation}\label{eq:eta_mean_1}
\la{\eta}\ra =  \frac{2}{Z^0 Z^\tau}\cosh(\beta_1 \nu^0 - \beta_2 \nu^\tau) \left(1-\frac{\nu^0}{\nu^\tau}\right) < \eta_\text{th}.
\end{equation}
It is always smaller than the thermodynamic efficiency $\eta_\text{th}$ (see Fig.~3). This can be understood by noting that generally $\la W/Q_2 \ra = \la W\ra /\la Q_2\ra - \text{cov}(W/Q_2,Q_2)/ \la Q_2\ra$, where $W= -(W_1+W_3)$ and $\text{cov}(W/Q_2,Q_2)$ denotes the covariance between the ratio $W/Q_2$ and $Q_2$ \cite{hei99}. The inequality $\la{\eta}\ra < \eta_\text{th}$ is then obeyed when stochastic efficiency and absorbed heat are positively correlated \cite{sup}. The high-temperature ($\beta_i \nu^j\ll 1$) and low-temperature ($\beta_i \nu^j\gg 1$) limits, $i=(1,2)$ and $j= (0,\tau)$, of the mean efficiency \eqref{eq:eta_mean_1} are readily evaluated. We obtain,
\begin{eqnarray}
	\langle \eta \rangle_\text{high} &=&  \frac{1}{2} \left(1-\frac{\nu^0}{\nu^\tau}\right)=\frac{\eta_\text{th}}{2},\\
	\langle \eta \rangle_\text{low} &=&\eta_\text{th} \left(e^{-2 \beta_2 \nu^\tau} + e^{-2 \beta_1 \nu^0} \right).
\end{eqnarray}
 The variance,  $\sigma_\eta^2 = \langle \eta^2 \rangle - \langle \eta \rangle^2$, of the stochastic efficiency may be evaluated in a similar manner, yielding,
 \be
 \label{21}
\sigma_\eta^2  = \frac{1}{4} \left( 1-\frac{\nu^0}{\nu^\tau} \right)^2
						    [1 - \tanh^2(\nu^0 \beta_1) \tanh^2(\nu^\tau \beta_2)].
\ee
 Its respective high and low temperatures limits are,
 \begin{eqnarray}
 {\sigma_\eta^2}_\text{high}&=&\frac{1}{4} \left( 1-\frac{\nu^0}{\nu^\tau} \right)^2= \frac{\eta_\text{th}^2}{4},\\
{\sigma_\eta^2}_\text{low} &=&\eta_\text{th}^2 \!
	\left( \frac{e^{2\beta_1 \nu^0} + e^{2 \beta_2 \nu^\tau} }{e^{2 \beta_1 \nu^0+2 \beta_2 \nu^\tau}+2e^{2  \beta_1 \nu^0}+2e^{2 \beta_2 \nu^\tau}} \right).
 \end{eqnarray}
 The behavior of both the average \eqref{eq:eta_mean_1} (inset) and the variance \eqref{21} of the  stochastic efficiency are represented as a function of inverse temperature in Fig.~4. The transition from a regime dominated by thermal fluctuations at high temperatures to a domain characterized by  quantum fluctuations at low temperatures is clearly visible. In particular, the mean efficiency sharply drops as the ratio $\text{cov}(W/Q_2,Q_2)/ \la Q_2\ra$ increases \cite{sup}, while the variance gets reduced when  thermal fluctuations are replaced by smaller quantum fluctuations.
\begin{figure}
	\centering
	\includegraphics[width=0.49\textwidth]{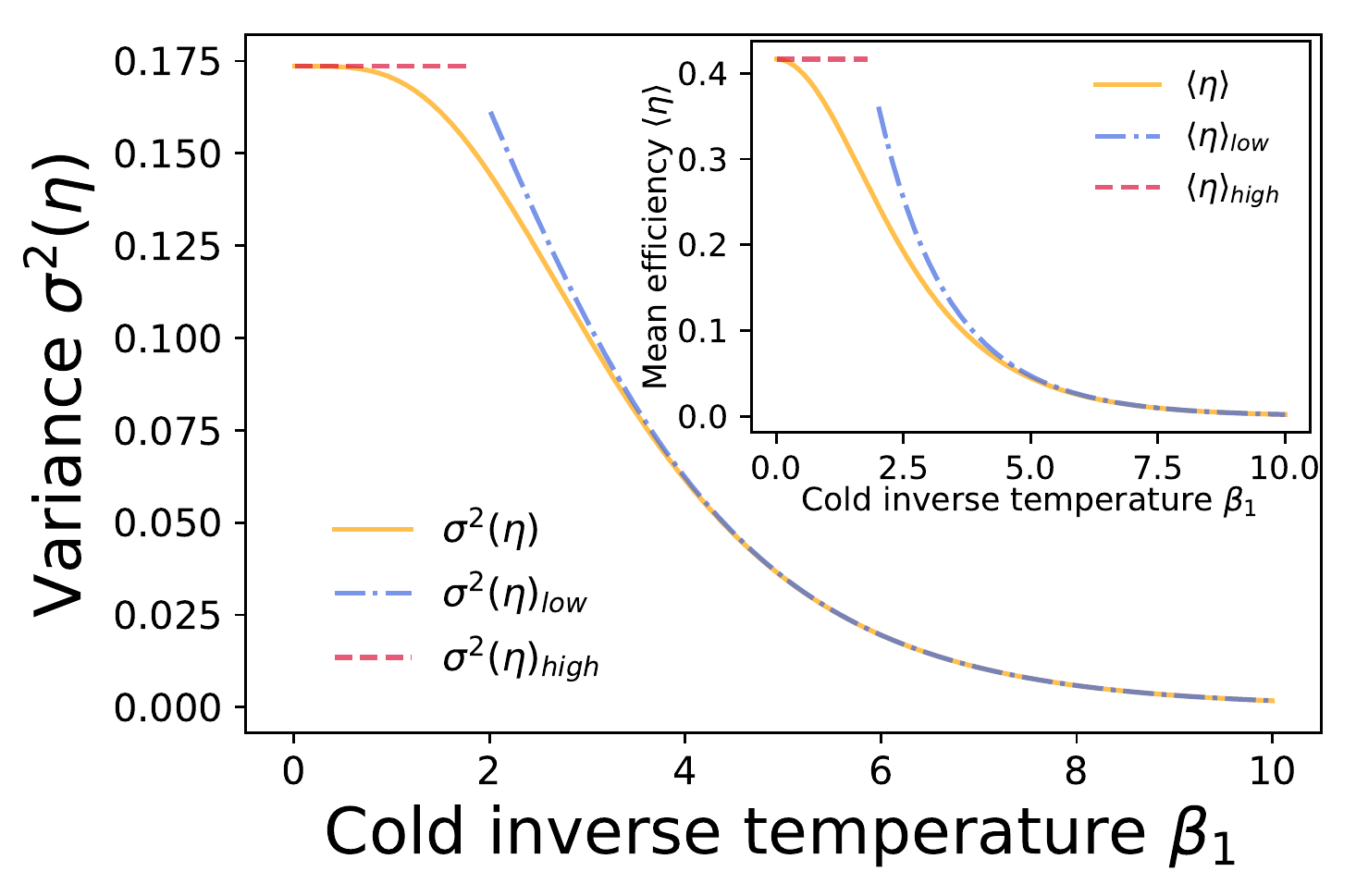}
	\caption{Average efficiency $\la \eta\ra$, Eq.~(18) (inset), and variance ${\sigma_\eta^2}$, Eq.~(21), as a function of the inverse cold temperature  $\beta_1=10\beta_2$ with the respective low-temperature (blue dotted-dashed) and high-temperature (red dotted-dashed) approximations (18)-(19) and (22)-(23). The crossover from a regime dominated by thermal fluctuations at high temperatures to a domain characterized by quantum fluctuations at low temperatures is clearly visible. Same parameters as in Fig.~2.}
\end{figure}

  \textit{Conclusions}.
We have developed a general framework allowing to  calculate the distribution of the efficiency of a quantum Otto engine. We have shown that it is fully determined by the time evolution operators of the two isothermal compression and expansion steps, and by the  two bath temperatures when complete thermalization is considered. The fluctuation statistics will additionally depend on the nonunitary relaxation dynamics in the case of incomplete thermalization. We have applied our results to an analytically solvable two-level engine and evaluated the discrete efficiency distribution in closed form. We have established the existence of peaks at infinity which follow from the quantum nature of the engine in the nonadiabatic regime. An average efficiency is thus not defined for nonadiabatic driving. 
We have additionally computed the first two cumulants of the stochastic efficiency in the adiabatic limit and found that the mean  is always smaller than the corresponding thermodynamic efficiency since efficiency and heat are positively correlated. We have finally observed the crossover of the variance from  the classical  to the quantum domain.

\textit{Acknowledgements.} We acknowledge financial support from the Volkswagen Foundation under project "Quantum coins and nano sensors" and the German Science Foundation (DFG) under project FOR 2724.

\subsection*{Supplemental Material}

\textit{Solution of the two-level model.} We here present the detailed solution of the expansion and compression dynamics of the spin-$1/2$ engine.	The Hamiltonian of a two-state system can be written in the general form,
	\begin{equation}
	\label{24}
		H = b_x(t) \sigma_x + b_y(t) \sigma_y + b_z(t) \sigma_z.
	\end{equation}
	In order to describe  a rotating magnetic field with frequency $\omega$ and time-dependent strength $\gamma(t)$, we set,
	\begin{equation}\label{eq:prefac}
		b_x(t)= \gamma(t) \cos(\omega t),~ b_y(t) =  \gamma(t) \sin(\omega t).
	\end{equation}
	We leave the function $b_z(t)$ unspecified for the time being.
A method to evaluate the corresponding time-evolution operator, given some constraints, has been proposed in Refs.~\cite{bar12,bar13}.
The first step is to write the time-evolution operator in the form,
 \begin{equation}\label{eq:unit}
U=
	\begin{pmatrix}
		u_{11} & -u_{21}^{*} \\
		u_{21} & u_{11}^{*}
	\end{pmatrix},
\end{equation}
where $u_{11}=\cos \chi \cdot e^{i \xi_{-} - i \phi/2}$ and $u_{21}= i \eta \sin \chi \cdot e^{i \xi_{+}+ \phi/2}$. Here $\eta= 1$ and the parameters $\xi_{\pm}$ are given by,
\begin{equation}\label{eq:pm}
	\xi_{\pm} = \int_0^t dt' ~ \sqrt{1 - \frac{\dot{\chi}^2}{\beta^2}} \csc(2 \chi) \pm \frac{1}{2} \sin^{-1}\left(\frac{\dot{\chi}}{\beta}\right) \pm \eta \frac{\pi}{4}.
\end{equation}
The given constraints allow the variables $\chi$, $\beta$, and $\phi$ to be chosen arbitrarily.
Using Eqs.~\eqref{eq:unit} and \eqref{eq:pm} the prefactors in the Hamilton operator  \eqref{24} are,
	\begin{eqnarray}
		b_x(t) &=& \beta \cos \phi, ~ b_y(t) = \beta \sin \phi, \\
		b_z(t) &=& \frac{\ddot{\chi}- \dot{\chi}\dot{\beta}/\beta}{2 \beta \sqrt{1-\dot{\chi}^2/\beta^2}} - \beta \sqrt{1-\dot{\chi}^2/\beta^2} \cot(2 \chi) + \frac{\dot{\phi}}{2}. \nonumber
	\end{eqnarray}
The choice \eqref{eq:prefac} is reproduced by taking $\beta = \gamma(t)$ and $\phi = \omega t$. By further setting $\chi = - \eta \int_0^t dt' \beta(t')$, we obtain $	b_z(t) = {\omega}/{2}$. The evolution operator \eqref{eq:unit}  then follows as,
\begin{equation}
	U=
	\begin{pmatrix}
		e^{-i \omega t/2} \cos I & i e^{- i \omega t /2} \sin I \\
		i e^{i \omega t/2} \sin I  & e^{i \omega t/2} \cos I
	\end{pmatrix},
\end{equation}
with the quantity $I = - \int_0^t dt' ~ \gamma(t')$.

\begin{figure}[t]
		\centering
		\includegraphics[width=0.49\textwidth]{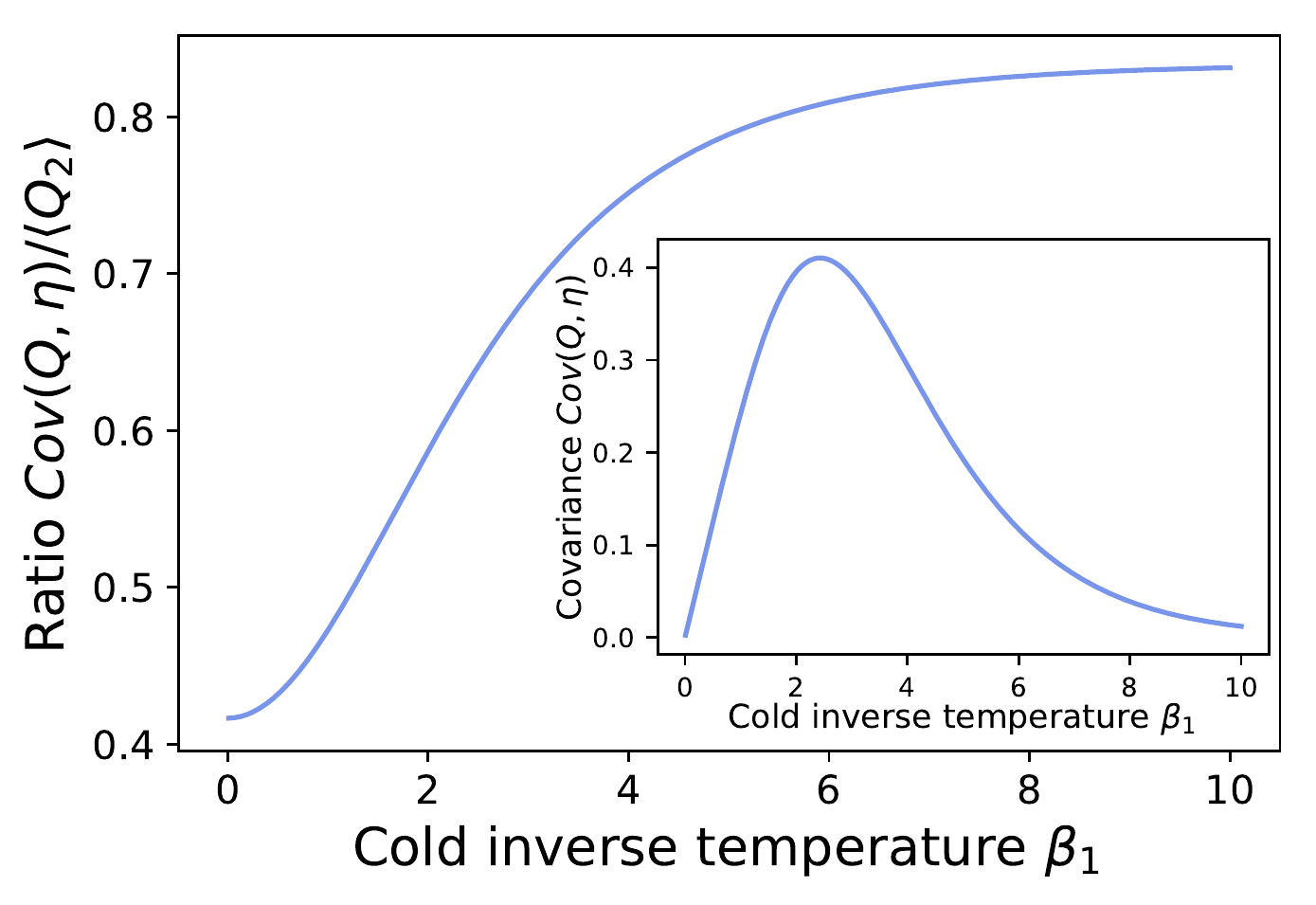}
		\caption{Covariance between absorbed heat $Q_2$ and stochastic efficiency $\eta$ for different cold inverse temperatures $\beta_1=10\beta_2$. Parameters are $u=1$,  $\gamma_0=0.5$ and $\gamma_\tau=3$.}\label{fig: appendix}
	\end{figure}

\textit{Zero-over-zero efficiency.} The computation of the efficiency distribution for the two-level engine, Eq.~(17) of the main text, involves expressions of the form $0/0$, which occur with finite probability. Since these are ill defined mathematically, we determine their values physically by concretely considering the case of adiabatic driving, $u=1$. The corresponding efficiency distribution is,
\bea
P(\eta) &=& P_{0}^0 P_{0}^\tau \delta\left(\eta-\frac{0}{0}\right) + P_{1}^0 P_{1}^\tau  \delta \left(\eta-\frac{0}{0}\right) \nonumber \\&+& \left(P_0^0 P_1^\tau + P_1^0 P_0^\tau \right) \delta\left[\eta-\left(1 - \frac{\nu^0}{\nu^\tau}\right)\right],
\eea
 since $|\bra{n} U_\text{exp}\ket{m}|^2=\delta_{nm}$ and $|\bra{k} U_\text{com}\ket{l}|^2=\delta_{kl}$. The first term, $P_{0}^0P_{0}^\tau=1$, corresponds to two baths with equal and vanishingly small temperatures, $\beta_1=\beta_2\rightarrow \infty$ (and the engine always remains in the ground state). Similarly, the second term, $P_{1}^0P_{1}^\tau=1$, corresponds to two baths with equal and extremely large temperatures, $\beta_1=\beta_2\rightarrow 0$ (and the engine always remains in the exciteted state). In both cases, the Carnot formula  implies that the efficiency vanishes and we therefore set $0/0=0$.

\textit{Calculation of the mean values.}
The average values of work and heat along the different branches of the heat engine cycle are obtained by direct integration of the corresponding probability distributions. We find,	
	\begin{eqnarray}
			\langle W_1 \rangle  &=& \int\limits_{- \infty}^{+\infty} dW_1 W_1 P(W_1) \\
			&=& \sum_{n,m} (\nu_m^\tau - \nu_n^0) \frac{e^{-\beta_1 \nu_n^0}}{Z^0} P_{n \rightarrow m}^\tau \\
			&=& [\nu^0 + \nu^\tau (1- 2u)] \tanh(\beta_1 \nu^0),
	\end{eqnarray}
and in an analogous manner,
\begin{equation}
	\langle W_3 \rangle = [\nu^\tau + \nu^0(1-2u)] \tanh(\beta_2 \nu^\tau).
\end{equation}
At the same time, the absorbed heat reads,
\begin{eqnarray}
		\langle Q_2 \rangle &=& \int\limits_{- \infty}^{+\infty} \int\limits_{- \infty}^{+\infty} dW_1 dQ_2 P(Q_2|W_1) p(W_1) \\
		&=& \sum_{k,m,n} (\nu_k^\tau - \nu_m^\tau) \frac{e^{- \beta_2 \nu_k^\tau - \beta_1 \nu_n^0}}{Z^\tau Z^0} P^\tau_{n \rightarrow m} \\
		&=& -\nu^\tau \left[ \tanh(\beta_2 \nu^\tau) + \tanh(\beta_1 \nu^0)(1 - 2u) \right].
\end{eqnarray}

We mention that averaged work and heat can also be calculated by considering the energy changes along individual branches of the Otto cycle, that is, by only performing projective energy measurements at the beginning and at the end of one given step, instead of the first three consecutive steps as done above. The two methods give the same results. This may be understood by noting that work and heat only depend on (diagonal) energy differences which do not depend on the (nondiagonal) coherences of the two-level  system. The presence or absence of intermediate projective energy measurements hence    do not affect the value of the averaged work and heat heat.

\textit{Covariance.} The covariance between  absorbed heat $Q_2$ and stochastic efficiency $\eta$ reads \cite{hei99},
\begin{equation}
	\text{Cov}(Q_2, \eta) = -(\langle W_1 \rangle + \langle W_3 \rangle)- \langle Q_2 \rangle \langle \eta \rangle.
\end{equation}
We have, as a result,
\begin{equation}
	\begin{aligned}
	\langle \eta \rangle &= \frac{-( \langle W_1 \rangle + \langle W_3 \rangle )}{\langle Q_2 \rangle} - \frac{\text{Cov}(Q_2,\eta)}{\langle Q_2 \rangle} \\
	&= \eta_\text{th} - \frac{\text{Cov}(Q_2,\eta)}{\langle Q_2 \rangle}
	\end{aligned}
\end{equation}
The covariance is shown in  Fig.~\ref{fig: appendix} as a function of the inverse temperature. We notice that it first increases and then decreases as the temperature is lowered. It is moreover always positive, indicating that heat and efficiency are always positively correlated. On the other hand, the ratio $\text{cov}(W/Q_2,Q_2)/ \la Q_2\ra$ increases. Consequently, the mean efficiency is smaller than the corresponding thermodynamic efficiency, $\langle \eta \rangle \leq \eta_\text{th}$.

\end{document}